\documentclass[aps, pra, showpacs, twocolumn, 
amssymb,superscriptaddress]{revtex4}

\usepackage{graphicx, bm, amsmath, amsfonts}

\begin{document}


\title{
On the existence of an energy gap in
one-dimensional Lesanovsky's model
}

\author{Hosho Katsura}
\affiliation{
Department of Physics, Gakushuin University, 
Mejiro, Toshima-ku, Tokyo 171-8588, Japan\\
}

\date{\today}
\begin{abstract}
We study the quantum lattice gas model in one dimension introduced by Lesanovsky~\cite{Lesanovsky_2012}, 
who showed that the exact ground state and a couple of excited states can be 
obtained analytically. The Hamiltonian of the model depends solely on the parameter 
$z$, the meaning of which is a fugacity in the corresponding classical lattice gas model.
For small $z$ ($0<z<1$), we prove that there is an energy gap between 
the ground state and the excited states by applying Knabe's method~\cite{Knabe_JSP}.
\end{abstract}


\maketitle
\section{Introduction}
Recently there has been a growing interest in exploring the physics of strongly interacting systems using Rydberg atoms, atoms in states of high principal quantum number $n$. It was argued that a variety of exotic systems can be simulated with Rydberg atoms in optical lattices
\cite{Weimer_PRL,Sela_PRB,Pohl_PRL,Olmos_PRA,Lesanovsky_2011,Lesanovsky_2012, Ates_Lesanovsky_2011, Lesanovsky_Katsura_PRA}. 
These atoms are interacting via the van der Waals-type interaction, which is strongly enhanced when $n$ is very large. This interaction naturally leads to the Rydberg blockade, i.e., a simultaneous excitation of two nearby atoms to Rydberg states is forbidden. The first experiments on Rydberg atoms in quasi-one-dimensional optical lattices were carried out~\cite{Viteau_PRL}. 

In Ref. \cite{Lesanovsky_2012} Lesanovsky has introduced a solvable model of Rydberg lattice gas in one dimension. The model can be thought of as a quantum Ising chain with long-range interaction in a transverse and longitudinal field. 
The model depends solely on the parameter $z$, the meaning of which is a fugacity in the corresponding classical lattice gas with hard-core constraint. 
Lesanovsky showed that the exact zero-energy ground state is a weighted superposition of states, each of which is labeled by a configuration of Rydberg states with the Rydberg blockade. 
He also obtained explicit expressions for a couple of excited states. 
The results of exact diagonalization suggest that it is likely that one of excited states he obtained corresponds to the first excited state. 
If this is the case, the energy gap is nonvanishing since the analytical expression for the energy of the first excited state is nonzero for finite $z$. 
To date, however, there is no rigorous proof of the existence of an energy gap. 

In this brief report, we show that the existence of an energy gap for small $z$ can be proved without knowing the explicit expression for the energy of the first excited state. 
The idea is to use the method proposed by Knabe in \cite{Knabe_JSP}, 
which was applied to show the existence of the gap in one-dimensional (1D) Affleck-Kennedy-Lieb-Tasaki models with various spins \cite{AKLT_CMP}. The method enables one to get lower bounds for energy gaps of infinite chains by diagonalizing finite-size chains with open boundaries. 
Other systems where this method is applicable include a sawtooth chain and spin ladders \cite{PRB_Sen_Shastry, PRB_Momoi_Hikihara}. 

\section{Lesanovsky's model}
We consider a system of hard-core bosons on a lattice. 
The Hilbert space at each lattice site is spanned by $|n_i \rangle$, 
where $n_i = 1$ ($0$) indicates that the site $i$ is occupied (empty). 
With the identification $|\uparrow \rangle \leftrightarrow |1 \rangle$ and 
$|\downarrow \rangle \leftrightarrow |0 \rangle$, the operator that creates/annihilates 
the hard-core boson at site $i$ can be expressed as $\sigma^\pm_i = (\sigma^x_i \pm i \sigma^y_i)/2$, 
where $\sigma^\alpha_i$ ($\alpha=x,y,z$) are the standard Pauli matrices. 
The model can be defined on any lattice in any dimension~\cite{PRA_TTK}. 
However, for simplicity, we restrict our attention to the case of 
1D lattice with periodic boundary conditions~\cite{Lesanovsky_2012}. 
An extension of a proof of the gap and exact results for the excited states 
in higher dimensional cases will be presented in a separate publication \cite{Katsura_prep}. 

The Hamiltonian for the 1D chain of length $N$ is given by
\begin{equation}
H = \sum^N_{i=1} P_{i-1} [\sigma^x_i + z P_i +z^{-1} n_i] P_{i+1}, 
\label{eq:ham1}
\end{equation}
where the parameter $z$ is real and positive. 
Here $P_i = (1-\sigma^z_i)/2$ and $n_i = (1+\sigma^z_i)/2$ are projectors 
on the occupied and empty states on the $i$-th site, respectively. 
The periodic boundary conditions imply $P_{N+1}=P_1$, $n_{N+1}=n_1$, and so on. 
In the following, we are interested in the restricted Hilbert space in which 
$n_i n_{i+1}=0$ for all $i$, i.e., an occupied state $|1\rangle$ 
is always accompanied by an empty state $|0\rangle$ on either side. 
This exclusion rule follows naturally from the Rydberg blockade. 
In the subspace, as shown in \cite{PRA_TTK}, the ground state of $H$ is unique~\cite{uniqueness}. 

\section{Knabe's method}

In this section, we provide a summary of Knabe's method~\cite{Knabe_JSP}. 
Let us consider a Hamiltonian of the form
\begin{equation}
{\cal H} = \sum^N_{i=1} {\cal Q}_i,
\end{equation}
with periodic boundary conditions (${\cal Q}_{N+1}={\cal Q}_1$). 
Here, ${\cal Q}_i$ ($i=1,2,...,N$) are projection operators, i.e., ${\cal Q}^2_i = {\cal Q}_i$, 
and their commutators satisfy $[{\cal Q}_i, {\cal Q}_j]=0$ if $|i-j|>1$~\cite{Rachel_PRB}. 
Note that these ${\cal Q}_i$'s take the same form as ${\cal Q}_1$, 
but act on different sites. 
The Hamiltonian is positive semi-definite by construction. 
If it is known that the ground-state energy is zero, then the inequality 
\begin{equation}
{\cal H}^2 \ge \epsilon {\cal H},~~~\epsilon>0
\end{equation}
implies that the energy gap (the lowest non-vanishing eigenvalue of ${\cal H}$) is larger than $\epsilon$ \cite{operator_ineq}. 

Knabe has shown that one can derive such an inequality if the same model 
on a finite chain with open boundaries satisfies
\begin{equation}
h^2_{n,i} \ge \epsilon_n h_{n,i},~~~\epsilon_n > \frac{1}{n},
\label{ineq1}
\end{equation}
where $n \ge 2$ and $h_{n,i} := \sum^{i+n-1}_{j=i}{\cal Q}_j$. 
Thus, the existence of the energy gap is established if one can show that 
the inequality (\ref{ineq1}) is satisfied for some integer $n$. 
Note that for such $n$, a lower bound for the energy gap is given by
\begin{equation}
\epsilon = \frac{n}{n-1} \left( \epsilon_n -\frac{1}{n} \right).
\label{eq:lower_bound}
\end{equation}

\section{Proof of the gap}

Let us apply Knabe's method to the Hamiltonian (\ref{eq:ham1}). 
For convenience we introduce a scaled Hamiltonian
\begin{eqnarray}
{\cal H} &=&  H/(z+z^{-1}) = \sum^N_{i=1} {\cal Q}_i, \\
{\cal Q}_i &=& \frac{1}{z+z^{-1}}P_{i-1} [\sigma^x_i + z P_i +z^{-1} n_i] P_{i+1}.
\end{eqnarray}
The condition that the ground-state energy is zero is satisfied
since the scaling of the Hamiltonian does not change the zero energy. 
The local Hamiltonians ${\cal Q}_i$ ($i=1,2,...,N$) 
are the projection operators that satisfy ${\cal Q}^2_i = {\cal Q}_i$, 
which can be verified using $\sigma^x_i P_i = n_i \sigma^x_i = \sigma^+_i$ 
and $P_i \sigma^x_i = \sigma^x_i n_i = \sigma^-_i$. 
It is also easy to see that $[{\cal Q}_i, {\cal Q}_j]=0$ if $|i-j|>1$. 

We first examine $h_{2,i} = {\cal Q}_i + {\cal Q}_{i+1}$. 
In the restricted Hilbert space, $h_{2,i}$ is expressed as the following matrix:
\begin{equation}
h_{2,i}=\frac{1}{z+z^{-1}}
\begin{pmatrix}
2z & 0 & 1 & 1 & 0 & 0 & 0 & 0 \\
0 & z & 0 & 0 & 0 & 1 & 0 & 0 \\
1 & 0 & z^{-1} & 0 & 0 & 0 & 0 & 0 \\
1 & 0 & 0 & z^{-1} & 0 & 0 & 0 & 0 \\
0 & 0 & 0 & 0 & z & 0 & 1 & 0 \\
0 & 1 & 0 & 0 & 0 & z^{-1} & 0 & 0 \\
0 & 0 & 0 & 0 & 1 & 0 & z^{-1} & 0 \\
0 & 0 & 0 & 0 & 0 & 0 & 0 & 0
\end{pmatrix}
\end{equation}
The order of the basis states is $|0000\rangle$, $|0001\rangle$, $|0010\rangle$, $|0100\rangle$, $|1000\rangle$, $|0101\rangle$, $|1010\rangle$, $|1001\rangle$. 
One can diagonalize $h_{2,i}$ analytically and find that the energy gap 
(the lowest non-vanishing eigenvalue of $h_{2,i}$) is given by
$\epsilon_2 = 1/(1+z^2)$ 
which is greater than $1/2$ when $0<z<1$. 
Therefore, from (\ref{ineq1}) and (\ref{eq:lower_bound}), 
the Hamiltonian $H$ has an energy gap if 
$0<z<1$
and a lower bound for the gap is $(1-z^2)/(1+z^2)$. 
It should be noted that, to our knowledge, this is the first example in which 
Knabe's method is successfully applied to a proof of the gap 
without the aid of numerical diagonalization. 

The condition $0<z<1$ is sufficient for the existence of the energy gap, 
but of course not optimal. An improved condition can be obtained 
by considering $h_{n,i}$ with $n>2$. 
For example, if we take $n=4$, we find that the energy gap of $h_{4,i}$ is 
$\epsilon_4 = x/(1+z^2)$ where $x$ is the smallest root of the cubic equation:
$x^3-(4+3z^2)x^2+(5+7z^2+2z^4)x-(2+4z^2+z^4) =0$.
The condition (\ref{ineq1}), i.e., $\epsilon_4 > 1/4$ yields $0<z<1.3263...$
A sufficient condition for the existence of the gap can be extended by 
diagonalizing the Hamiltonians for longer chains with open boundaries. 

\section*{Acknowledgment}
The author thanks Igor Lesanovsky for valuable discussions. 
This work was supported by Grant-in-Aid 
for Young Scientists (B) (23740298). 



\begin{thebibliography}{99}
\bibitem{uniqueness}
The uniqueness of the ground state in the same model on an arbitrary lattice 
was shown in \cite{PRA_TTK}. Note that the signs of the off-diagonal elements of $H$ 
can be changed by means of the unitary transformation $H \to \Sigma^z H \Sigma^z$, 
where $\Sigma^z := \prod^N_{i=1} \sigma^z_i$.
\bibitem{operator_ineq}
Here we write ${\cal A} \ge {\cal B}$ to denote that ${\cal A}-{\cal B}$ is positive semi-definite.
\end{thebibliography}
\end{document}